\documentclass{elsart}
\journal{Nuclear Physics A}
\usepackage{amssymb}
\usepackage[german,english]{babel}
\usepackage[intlimits,sumlimits,namelimits]{amsmath}
\usepackage[dvips]{epsfig}
\usepackage[dvips]{color}
\usepackage{amssymb,ifthen,shadow}
\usepackage{fancyhdr}
\usepackage[dvips]{epsfig}

\newcommand{\beq}{\begin{equation}}
\newcommand{\eeq}{\end{equation}}
\newcommand{\beqa}{\begin{eqnarray}}
\newcommand{\eeqa}{\end{eqnarray}}
\newcommand{\bra}[1]{\mbox{$\langle #1|$}}
\newcommand{\ket}[1]{\mbox{$|#1\rangle$}}

\begin{document}

\begin{frontmatter}

\title{\bf Final-state $NN$-rescattering in spin asymmetries of 
  $d(\gamma,\pi^-)pp$ reaction}

\author{Eed M.\ Darwish\corauthref{eed}}
\corauth[eed]{Corresponding author.\\
  {\it Electronic address:} eeddarwish@yahoo.com (E.M.\ Darwish).}
\address{Physics Department, Faculty of Science,
South Valley University,\\ Sohag 82524, Egypt}

\author{Agus Salam}
\address{Departemen Fisika, FMIPA, Universitas Indonesia, Depok 16424, 
Indonesia}  

\date{\today}

\begin{abstract}
  The role of the final-state $NN$-rescattering ($NN$-FSI) in the
  polarization observables of the inclusive reaction
  $d(\gamma,\pi^-)pp$, involving polarization of the photon beam
  and/or the deuteron target, is investigated. Various single- and
  double-spin asymmetries are studied with respect to the influence of
  such interaction effect and numerical predictions are given for
  forthcoming experiments. It has been found that the effect of
  $NN$-FSI is quite important for the single-spin asymmetries
  $\Sigma$, $T_{20}$, $T_{21}$ and $T_{22}$ and the double-spin
  asymmetry $T^{\ell}_{20}$, whereas it is much less important for the
  vector target asymmetry $T_{11}$ and other beam-target
  double-polarization asymmetries.  Furthermore, we found that the
  inclusion of the $NN$-FSI improves the description of the LEGS data
  for the linear photon asymmetry.

\vspace{0.2cm}

\end{abstract}
\end{frontmatter}
%%%%%%%%%%%%%%%%%%%%%%%%%%%%%%%%%%%%%%
\section{Introduction and Motivation} 
\label{sec1}
%%%%%%%%%%%%%%%%%%%%%%%%%%%%%%%%%%%%%%
A still very interesting topic in medium-energy nuclear physics is
concerned with the role of effective degrees of freedom in hadronic
systems in terms of nucleons, mesons and isobars. In particular, the
connection of such effective degrees of freedom to the underlying
quark-gluon dynamics of Quantum Chromodynamics (QCD) has still to be
clarified. For the study of these basic questions, the two-nucleon
system provides an important test laboratory. First of all, it is the
simplest nuclear system for the study of the $NN$ interaction.
Moreover, due to the lack of free neutron targets, the deuteron is the
simplest nucleus which can be used to extract neutron properties.

Interest in pion photoproduction from light nuclei has increased
mainly through the construction of new high-duty continuous electron
beam machines such as MAMI in Mainz, ELSA in Bonn, LEGS in Brookhaven
or JLab in Newport News (for an experimental overview see
\cite{Bur04,Kru03}). In conjunction with these experimental efforts,
we investigate in this work the reaction $d(\gamma,\pi^-)pp$ including
final-state interaction (FSI) effects with special emphasis on
polarization observables. Spin physics proves to be a vital element
and indispensable tool in our venture of revealing the internal
structure of hadronic matter. It is a well-known fact that in order to
study further details of a reaction one has to study polarization
observables like, e.g., beam asymmetry, target asymmetries and
beam-target asymmetries. These additional observables contain
interference terms of the various amplitudes in different combinations
from which one can determine all amplitudes provided one has measured
a complete set of observables.

Up to present times, most of calculations for incoherent pion
photoproduction from the deuteron have considered only the unpolarized
observables like differential and total cross sections
\cite{ChL51,LaF52,BlL77,Lag78,Lag81,ScA96,Lev01,Dar03,Faes02}.
Notwithstanding this continuing effort to study this process, the
wealth of information contained in it has not yet been fully
exploited. Since the $t$-matrix has 12 independent complex amplitudes,
one has to measure 23 independent observables, in particular, in order
to determine completely the $t$-matrix. The unpolarized differential
and total cross sections provide information only on the sum of the
absolute squares of the amplitudes, whereas polarization degrees of
freedom lead to new observables which provide additional information
on the nuclear structure as well as on the reaction mechanism.

Polarization observables for incoherent pion photoproduction on the
deuteron with polarized photon beam and/or polarized deuteron target
have been poorly investigated.  The influence of $NN$- and $\pi
N$-rescattering on the analyzing powers in the quasifree
$\pi^-$-photoproduction reaction on the deuteron in the
$\Delta$-resonance region has been studied within a diagrammatic
approach \cite{Log00} using relativistic-invariant forms of the
photoproduction and $\pi N$ scattering operators. In that work,
calculations for analyzing powers connected to beam and target
polarization, and to polarization of one of the final protons are
given. It has been shown that the effects of rescattering play a
noticeable role in the behaviour of these observables in the kinematic
region of the $\Delta$-isobar with large momenta of protons in the
final state. The deuteron tensor analyzing power components in
negative pion photoproduction from the deuteron have been studied in
the pure impulse approximation (IA) \cite{Log04} using deuteron wave
function of different realistic potential models. It was shown that
the influence of rescattering effects is necessary and must be taken
into account in the analysis of experimental data. In our previous
evaluation \cite{Dar03+}, special emphasizes are given for the
beam-target spin asymmetry of the total cross section and the
Gerasimov-Drell-Hearn (GDH) sum rule for the deuteron including $NN$-
and $\pi N$-rescattering in the final state. We found that FSI reduce 
the value of the GDH integral to about half of the value obtained for
the pure IA.

Various single- and double-polarization asymmetries of the
differential cross section in incoherent pion photoproduction from the
deuteron have been discussed in our recent papers
\cite{Dar04,Dar04E13,Dar04JG} without any kind of FSI effects. The
sensitivity of these spin observables to the model deuteron wave
function has been investigated. We found that interference of Born
terms and the $\Delta$(1232)-contribution plays a significant role.
Furthermore, our results for the linear photon asymmetry
\cite{Dar04JG,Dar05PLB} have been compared with the preliminary
experimental data from the LEGS Spin collaboration \cite{LEGS} and
major discrepancies were evident. Since a strong influence of the FSI
on the unpolarized cross sections was found in \cite{Dar03}, one might
expect that the LEGS data can be understood in terms of the FSI and
possibly two-body effects. It has been found in \cite{Dar05PLB}, that
$NN$-FSI is quite important and leads to a better agreement with
existing experimental data. Moreover, we have investigated most
recently the influence of final-state $NN$-rescattering on the
helicity structure of the $\vec\gamma\vec d\to\pi^-pp$ reaction
\cite{Dar04NPA,Dar04PTP}. The differential polarized cross-section
difference for the parallel and antiparallel helicity states has been
predicted and compared with recent experimental data from MAMI
(Mainz/Pavia) \cite{Pedroni}.  It has been shown that the effect of
$NN$-rescattering is much less important in the polarized differential
cross-section difference than in the unpolarized one.

As a further step in this direction, we investigate in this paper the
role of the $NN$-rescattering effect in several single- and
double-polarization observables of photon and deuteron target in
negative pion photoproduction from the deuteron with polarized photon
beam and/or oriented deuteron target. The understanding of this
mechanism is of great importance to understand the basic $NN$
interaction. The $\pi N$-rescattering contribution is found to be
negligible \cite{Lev01,Dar03} and thus it is not considered in the
present work.  The second point of interest is to analyze the
preliminary experimental data for the linear photon asymmetry $\Sigma$
from LEGS \cite{LEGS} in order to keep up with the development of the
experimental side.

The structure of this paper is as follows. In Sect.\ \ref{sec2}, the
model for the elementary $\gamma N\to\pi N$ and $NN\to NN$ reactions
which will serve as an input for our calculation of the reaction on
the deuteron is briefly summarized. In Sect.\ \ref{sec3} we provide
some details about the reaction on the deuteron. The general formalism
and the separate contributions of the pure IA and the $NN$-FSI to the
transition matrix, based on time-ordered perturbation theory, are
described in this section. The formal expressions for various
polarization observables in terms of the transition amplitude are
given in Sect.\ \ref{sec4}. Details of the actual calculations and the
results for the pure IA as well as for the inclusion of the FSI in the
$NN$ system are presented and discussed in Sect.\ \ref{sec5}. Finally,
a summary and conclusions are given in Sect.\ \ref{sec6}.
%%%%%%%%%%%%%%%%%%%%%%%%%%%%%%%%%%%%%%%%%%%
\section{Elementary reactions}\label{sec2} 
%%%%%%%%%%%%%%%%%%%%%%%%%%%%%%%%%%%%%%%%%%%
Incoherent pion photoproduction from the deuteron is governed by basic
two-body processes, namely pion photoproduction on a nucleon and
hadronic two-body scattering reactions. For the latter only $NN$
scattering in the two-body subsystems is considered in this work. In
the following, we will briefly summarize these two elementary
processes, i.e., the $\gamma N\to\pi N$ and $NN\to NN$ reactions.
%%%%%%%%%%%%%%%%%%%%%%%%%%%%%%%%%%%%%%%%%%%%%%%%%%%%%%%%%
\subsection{The $\gamma N\to\pi N$ process}\label{sec21} 
%%%%%%%%%%%%%%%%%%%%%%%%%%%%%%%%%%%%%%%%%%%%%%%%%%%%%%%%%
The starting point of the construction of an operator for pion
photoproduction on the two-nucleon space is the elementary pion
photoproduction operator on a single nucleon, i.e., $\gamma N\to\pi
N$. In the present work we will examine the various observables for
the reaction on the nucleon using, as in our previous work
\cite{Dar03}, the effective Lagrangian model developed by Schmidt {\it
  et al.}  \cite{ScA96}. The main advantage of this model is that it
has been constructed in an arbitrary frame of reference and allows a
well defined off-shell continuation as required for studying pion
production on nuclei. It consists of the standard pseudovector Born
terms and the contribution of the $\Delta(1232)$-resonance. For
further details with respect to the elementary pion photoproduction
operator we refer to \cite{ScA96}. As shown in Figs.\ 1-3 in our
previous work \cite{Dar03}, the results of our calculations for the
elementary process are in good agreement with recent experimental data
as well as with other theoretical predictions and gave a clear
indication that this elementary operator is quite satisfactory for our
purpose, namely to incorporate it into the reaction on the deuteron.
%%%%%%%%%%%%%%%%%%%%%%%%%%%%%%%%%%%%%%%%%%%%%%%%%%%%%
\subsection{Nucleon-nucleon scattering}\label{sec22} 
%%%%%%%%%%%%%%%%%%%%%%%%%%%%%%%%%%%%%%%%%%%%%%%%%%%%%
For the nucleon-nucleon scattering in the $NN$-subsystem we use in
this work a specific class of separable potentials \cite{HaP8485}
which historically have played and still play a major role in the
development of few-body physics and also fit the phase shift data for
$NN$-scattering. The EST method \cite{Ern7374} for constructing
separable representations of modern $NN$ potentials has been applied
by the Graz group \cite{HaP8485} to cast the Paris potential
\cite{La+80} in separable form. This separable model is most widely
used in case of the $\pi NN$ system (see for example \cite{Gar90} and
references therein). Therefore, for the present study of the influence
of $NN$-rescattering this model is good enough.
%%%%%%%%%%%%%%%%%%%%%%%%%%%%%%%%%%%%%%%%%%%%%%%%%%%%%%%
\section{The $\gamma d\to\pi NN$ reaction}\label{sec3}
%%%%%%%%%%%%%%%%%%%%%%%%%%%%%%%%%%%%%%%%%%%%%%%%%%%%%%%
Following the formulation of our previous work \cite{Dar03}, the
general form for the unpolarized cross section in incoherent pion
photoproduction from the deuteron is given according to \cite{BjD64}
by
\begin{eqnarray}
d\sigma
&=&
\frac{\delta^{4}(k+d-p_1-p_2-q)M^2_{N}d^{3} p_1d^{3} p_2d^{3} q}
{96(2 \pi)^{5} |\vec v_{\gamma} - \vec
v_{d}|\omega_{\gamma}E_{d}E_1E_2\omega_q} 
\nonumber \\
& & \times~ 
\sum_{smt} \,\sum_{m_{\gamma}m_d} 
\left |{\mathcal M}^{(t\,\mu)}_{s\,m\,m_{\gamma}\,m_d}
(\vec p_1,\vec p_2,\vec q,\vec k,\vec d) \right|^{2}\,,
\label{eq:1}
\end{eqnarray}
The semi-inclusive
unpolarized differential cross section, where only the final pion is
detected without analyzing its energy, is given from (\ref{eq:1}) by
\begin{eqnarray}
\label{eq:2}
\frac{d\sigma}{d\Omega_{\pi}} & = & \frac{1}{6}~\int_0^{q_{max}}dq\int\hspace{-0.2cm}\int d\Omega_{p_{NN}}\, 
\rho_{s}\, 
\sum_{smt} \,\sum_{m_{\gamma}m_{d}}\left \vert {\mathcal
M}^{(t\,\mu)}_{s\,m\,m_{\gamma}\,m_{d}}(\vec{p}_{NN},\vec q,\vec k) 
\right \vert^{2}\,.
\end{eqnarray}

The transition ${\mathcal M}$-matrix elements are calculated in the 
frame of time-ordered perturbation theory, using the elementary pion 
photoproduction operator introduced in section \ref{sec2} and including 
$NN$-rescattering in the final state. The amplitude of the 
$\gamma d\to\pi NN$ reaction is determined by the transition matrix 
\begin{eqnarray}
\label{eq:4}
{\mathcal M}^{(t\mu)}_{sm
m_{\gamma}m_d}(\vec{k},\vec{q},\vec{p_1},\vec{p_2}) 
& = &
^{(-)}\bra{\vec{q}\,\mu,\vec{p_1}\vec{p_2}\,s\,m\,t-\mu}\epsilon_{\mu}  
(m_{\gamma})J^{\mu}(0)\ket{\vec{d}\,m_d\,00}\,,
\end{eqnarray}
where $J^{\mu}(0)$ denotes the current operator. In principle, the
full treatment of all interaction effects requires a  
three-body treatment. However, the outgoing $\pi NN$ scattering state is 
approximated in this work by  
\begin{eqnarray}
\ket{\vec{q}\,\mu,\vec{p_1}\vec{p_2}\,s\,m\,t-\mu}^{(-)} &=&
\ket{\vec{q}\,\mu,\vec{p_1}\vec{p_2}\,s\,m\,t-\mu} 
\nonumber \\
& & 
+ \hat G_{0}^{\pi NN (-)}
\,\hat T^{NN} \ket{\vec{q}\,\mu,\vec{p_1}\vec{p_2}\,s\,m\,t-\mu}\,,
\label{eq:5}
\end{eqnarray}
%%%%%%%%%%%%%%%%
\begin{figure}[htb]
\hspace*{1cm}\includegraphics[scale=0.8]{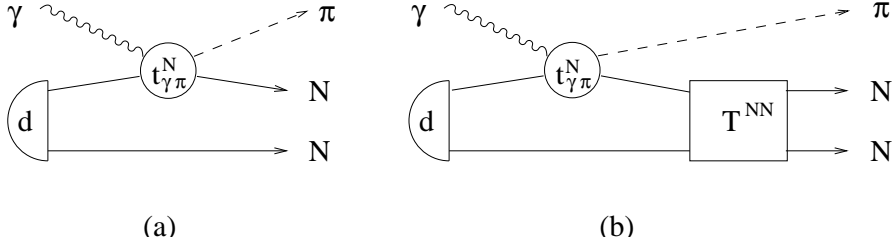}
\caption{Diagrammatic representation of incoherent pion photoproduction 
  from the deuteron including $NN$-rescattering in the final state: 
  (a) impulse approximation (IA) and (b) $NN$-rescattering.} 
\label{t-matrix}
\end{figure}
%%%%%%%%%%%%%%%%
where $\ket{\vec{q}\,\mu,\vec{p_1}\vec{p_2}\,s\,m\,t-\mu}$ denotes the 
free $\pi NN$ plane wave, $\hat G_{0}^{\pi NN (-)}$ the free $\pi NN$
propagator and $\hat T^{NN}$ the reaction operator for $NN$-scattering. 
In this approximation, the total transition matrix element reads  
\begin{eqnarray}
\label{eq:6}
{\mathcal M}^{(t\mu)}_{sm m_{\gamma}m_d} & = &
{\mathcal M}_{sm m_{\gamma}m_d}^{(t\mu)~IA} + 
{\mathcal M}_{sm m_{\gamma}m_d}^{(t\mu)~NN}\,,
\end{eqnarray}
where the first term denotes the transition matrix of the pure IA
which means that the reaction will take place only on one of the
nucleons leaving the other as a pure spectator (see diagram (a) in
Fig.\ \ref{t-matrix}). The second term represents the corresponding
matrix for the $NN$-rescattering (see Fig.\ \ref{t-matrix}(b)).

Denoting the matrix element of the elementary production on a nucleon
by $\hat t_{\gamma\pi}^{N}$, the amplitude of the IA-term in the
deuteron lab-frame reads \cite{Dar03}
\begin{eqnarray}
\label{eq:7}
  {\mathcal M}_{sm m_{\gamma}m_d}^{(t\mu)~IA}
  (\vec k,\vec q,\vec p_1,\vec p_2) &=&
%\bra{\vec{p_1}\vec{p_2}\,s\,m\,t-\mu}
%\hat{{\mathcal M}}^{IA}(\vec{k},m_{\gamma},\vec{q},\mu)
%\ket{\vec{d}=0\,m_d\,,00}\nonumber \\
%&=& 
\sqrt{2}\sum_{m^{\prime}}\langle s 
  m,\,t -\mu|\,\Big( \langle
  \vec{p}_{1}|\hat t_{\gamma\pi}^{N}(\vec k,\vec q\,)|-\vec{p}_{2}\rangle
  \tilde{\Psi}_{m^{\prime},m_{d}}(\vec{p}_{2}) 
\nonumber\\ & &  \hspace{1cm} 
-(-)^{s+t}(\vec p_1 \leftrightarrow \vec p_2) 
\Big)\,|1
  m^{\prime},\,00\rangle \,,
\end{eqnarray}
where $|1 m^{\prime},\,00\rangle$ denotes the two-nucleon spin and 
isospin wave function.

By using the information in Ref.\ \cite{Dar03}, we can write the 
$NN$-rescattering term in (\ref{eq:6}) as 
\begin{eqnarray}
\label{eq:9}
{\mathcal M}^{(t\mu)~NN}_{sm m_{\gamma}m_d}
(\vec k,\vec q,\vec p_1,\vec p_2) &=&
%\bra{\vec{q}\,\mu,\vec{p_1}\vec{p_2}\,s\,m\,t-\mu}
%\hat{{\mathcal M}}^{NN}(\vec k,m_{\gamma})
%\ket{\vec{d}=0\,m_d\,,00}\nonumber \\
%& = & 
\sum_{m^{\prime}}\int\hspace{-0.2cm}\int\hspace{-0.2cm}\int d^3
\vec p^{\,\prime}_{NN} 
\sqrt{\frac{E_1 E_2}{E_1' E_2'}}
\,\widetilde {\mathcal R}_{s m m^{\prime}}^{NN,\,t\mu}(W_{NN},\vec p_{NN},\vec
p^{\,\prime}_{NN}) 
\nonumber\\ & & \times ~
\frac{M_N}{\widetilde p^{\, 2} - p_{NN}^{\prime\,2} + i\epsilon}
{\mathcal M}^{(t\mu)~IA}_{sm^{\prime},m_{\gamma}m_d}(\vec k,\vec
q,\vec p^{\,\prime}_1,\vec p^{\,\prime}_2)\,.
%\nonumber\\ &&
\end{eqnarray}
The conventional $NN$-scattering matrix $\widetilde {\mathcal 
R}_{smm^{\prime}}^{NN,\,t\mu}$ is introduced with respect to
noncovariantly normalized states. It is expanded in terms of the
partial wave contributions ${\mathcal
T}_{Js\ell\ell^{\prime}}^{NN,\,t\mu}$  as follows 
\begin{eqnarray}
\label{eq:10}
\widetilde {\mathcal R}_{smm^{\prime}}^{NN,\,t\mu}(W_{NN},\vec p_{NN},\vec
 p^{\,\prime}_{NN}) 
 & = & \sum_{J}\,\sum_{\ell\ell^{\prime}}
 {\mathcal F}_{\ell\ell^{\prime}\,mm'}^{NN,\,Js}
 (\hat{p}_{NN},\hat{p}_{NN}^{\,\prime}) 
\nonumber \\
 & & \times~
{\mathcal T}_{Js\ell\ell^{\prime}}^{NN,\,t\mu}
 (W_{NN},p_{NN},p_{NN}^{\,\prime})\, , 
\end{eqnarray}
where the purely angular function ${\mathcal
F}_{\ell\ell^{\prime} mm^{\prime}}^{NN,\,Js}
(\hat{p}_{NN},\hat{p}_{NN}^{\,\prime})$ is defined by  
\begin{eqnarray}
\label{eq:11}
{\mathcal F}_{\ell\ell^{\prime} mm^{\prime}}^{NN,\,Js}(\hat{p}_{NN},
\hat{p}_{NN}^{\,\prime}) & = & 
 \sum_{M}\,\sum_{m_{\ell}m_{\ell^{\prime}}} C^{\ell s J}_{m_{\ell} m M}\,
C^{\ell^{\prime} s J}_{m_{\ell^{\prime}} m^{\prime} M}
 Y^{\star}_{\ell m_{\ell}}(\hat{p}_{NN})
 Y_{\ell^{\prime} m_{\ell^{\prime}}}(\hat{p}_{NN}^{\,\prime})\,.
\end{eqnarray}
The necessary half-off-shell $NN$-scattering matrix ${\mathcal
T}_{Js\ell\ell^{\prime}}^{NN,\,t\mu}$ was obtained from separable
representation of a realistic $NN$-interaction \cite{HaP8485}.
Explicitly, all $S$, $P$, and $D$ waves were included in the
$NN$-scattering matrix.
%%%%%%%%%%%%%%%%%%%%%%%%%%%%%%%%%%%%%%%%%%%%%%%%%%%%%%%%%%%%
\section{Polarization observables}
\label{sec4}
%%%%%%%%%%%%%%%%%%%%%%%%%%%%%%%%%%%%%%%%%%%%%%%%%%%%%%%%%%%%
Since calculational details associated with the evaluation of the
polarization observables in incoherent pion photoproduction from the
deuteron were considered in our previous papers \cite{Dar04,Dar04JG}
there is no need to repeat them here. However, we briefly recall the
necessary definitions of observables.

In the present work, we consider the following polarization observables:\\
(i) The linear photon asymmetry
\beqa
\label{eq:17}
\Sigma & = & \frac{2}{\mathcal Q} \,{\mathcal R}e\,\sum_{smt}\sum_{m_d}
\int_{0}^{q_{\rm max}}dq\int\hspace{-0.2cm}\int d\Omega_{p_{NN}} \,\rho_s\, 
{\mathcal M}^{(t\mu)}_{sm+1m_d}\,({\mathcal M}^{(t\mu)}_{sm-1m_d})^{\star}\,,
\eeqa
where
\begin{eqnarray}
\label{eq:3}
{\mathcal Q} & = & \int_0^{q_{max}}dq\int\hspace{-0.2cm}\int d\Omega_{p_{NN}}\, 
\rho_{s}\, 
\sum_{smt} \,\sum_{m_{\gamma}m_{d}}\left \vert {\mathcal
M}^{(t\,\mu)}_{s\,m\,m_{\gamma}\,m_{d}}(\vec{p}_{NN},\vec q,\vec k) 
\right \vert^{2}\,.
\end{eqnarray}
(ii) The vector target asymmetry
\beqa
\label{eq:19}
T_{11} & = & \frac{\sqrt{6}}{\mathcal Q} \,{\mathcal I}m\,\sum_{smt}
\sum_{m_{\gamma}}
\int_{0}^{q_{\rm max}}dq\int\hspace{-0.2cm}\int d\Omega_{p_{NN}} \,\rho_s\, 
\Big[{\mathcal M}^{(t\mu)}_{smm_{\gamma}-1}-
{\mathcal M}^{(t\mu)}_{smm_{\gamma}+1}\Big] 
\nonumber \\ & & \times ~ 
({\mathcal M}^{(t\mu)}_{smm_{\gamma}0})^{\star}\,.
\eeqa
(iii) The tensor target asymmetries
\beqa
\label{eq:21}
T_{20} & = & \frac{1}{\sqrt{2}{\mathcal Q}} \sum_{smt}
\sum_{m_{\gamma}}\int_{0}^{q_{\rm max}}dq\int\hspace{-0.2cm}\int 
d\Omega_{p_{NN}} \,\rho_s\, 
\Big[|{\mathcal M}^{(t\mu)}_{smm_{\gamma}+1}|^2 +
|{\mathcal M}^{(t\mu)}_{smm_{\gamma}-1}|^2 
\nonumber \\ & &  
- 2\,|{\mathcal M}^{(t\mu)}_{smm_{\gamma}0}|^2\Big]\,, 
\eeqa
\beqa
T_{21} & = & \frac{\sqrt{6}}{\mathcal Q} {\mathcal R}e\,\sum_{smt}
\sum_{m_{\gamma}}\int_{0}^{q_{\rm max}}dq\int\hspace{-0.2cm}\int 
d\Omega_{p_{NN}} \,\rho_s\, 
\Big[{\mathcal M}^{(t\mu)}_{smm_{\gamma}-1} - 
{\mathcal M}^{(t\mu)}_{smm_{\gamma}+1}\Big] 
\nonumber \\ & & \times~
({\mathcal M}^{(t\mu)}_{smm_{\gamma}0})^{\star}\,, 
\label{eq:22}
\eeqa
\beqa
T_{22} & = & \frac{2\sqrt{3}}{\mathcal Q} {\mathcal R}e\,\sum_{smt}
\sum_{m_{\gamma}}\int_{0}^{q_{\rm max}}dq\int\hspace{-0.2cm}\int 
d\Omega_{p_{NN}} \,\rho_s\, 
{\mathcal M}^{(t\mu)}_{smm_{\gamma}-1}
({\mathcal M}^{(t\mu)}_{smm_{\gamma}+1})^{\star}\,.
\label{eq:23}
\eeqa
(iv) The longitudinal photon and deuteron double-polarization asymmetries
\beqa
T^{\ell}_{20} & = & \frac{-1}{\sqrt{2}{\mathcal Q}} \sum_{smt}
\sum_{m_{\gamma}}\int_{0}^{q_{\rm max}}dq\int\hspace{-0.2cm}\int 
d\Omega_{p_{NN}} \,\rho_s\, 
\Big[({\mathcal M}^{(t\mu)}_{smm_{\gamma}-1})^{\star}
{\mathcal M}^{(t\mu)}_{s-mm_{\gamma}+1}
\nonumber \\ & & 
+ ({\mathcal M}^{(t\mu)}_{smm_{\gamma}+1})^{\star}
{\mathcal M}^{(t\mu)}_{s-mm_{\gamma}-1} 
%\nonumber \\ & & 
- 2\,({\mathcal M}^{(t\mu)}_{smm_{\gamma}0})^{\star}
{\mathcal M}^{(t\mu)}_{s-mm_{\gamma}0}\Big]\,, 
\label{eq:25}
\eeqa
\beqa
T^{\ell}_{2\pm 2} & = & \frac{-\sqrt{3}}{\mathcal Q}\sum_{smt} 
\sum_{m_{\gamma}}\int_{0}^{q_{\rm max}}dq\int\hspace{-0.2cm}\int 
d\Omega_{p_{NN}} \,\rho_s\,
({\mathcal M}^{(t\mu)}_{smm_{\gamma}\pm 1})^{\star}
{\mathcal M}^{(t\mu)}_{s-mm_{\gamma}\pm 1}\,.
\label{eq:26}
\eeqa

As shown in our previous work \cite{Dar04E13,Dar04JG}, the asymmetries
$T_{10}^c$, $T_{11}^c$, $T^c_{20}$, $T^c_{21}$, $T^c_{22}$
$T_{10}^{\ell}$, $T_{1\pm 1}^{\ell}$ and $T_{2\pm 1}^{\ell}$ vanish.
Therefore, in what follows we shall discuss the results for only the
$\Sigma$, $T_{11}$, $T_{20}$, $T_{21}$, $T_{22}$, $T_{20}^{\ell}$ and
$T_{2\pm 2}^{\ell}$ asymmetries.
%%%%%%%%%%%%%%%%%%%%%%%%%%%%%%%%%%%
\section{Discussion of results}
\label{sec5}
%%%%%%%%%%%%%%%%%%%%%%%%%%%%%%%%%%%
In this section we present our predictions of the polarization
observables defined in Sect.\ \ref{sec4} for the inclusive negative
pion photoproduction reaction from the deuteron. For the elementary
pion photoproduction operator on the free nucleon, the effective
Lagrangian model developed by Schmidt {\it et al.} \cite{ScA96} has
been considered. The two contributions to the pion production
amplitude on the deuteron, i.e., the IA in (\ref{eq:7}) and the
$NN$-rescattering in (\ref{eq:9}) are evaluated by taking a realistic
$NN$ potential model for the deuteron wave function and the $NN$
scattering amplitudes, in this work the Paris potential.

The discussion of the results is divided into three parts. First, we
discuss the influence of $NN$-FSI on the single-polarization
observables $\Sigma$, $T_{11}$, $T_{20}$, $T_{21}$, and $T_{22}$ by
comparing the pure IA with the inclusion of $NN$-rescattering in the
final state. In the second part, we consider the double-polarization
asymmetries for photon and deuteron target. In the third one, we
compare our results with the available experimental data.  In all the
plots that follow, the solid curves show the results of the full
calculation, i.e., when $NN$-rescattering is included, while the
dashed curves show the contribution of the pure IA alone in order to
clarify the importance of $NN$-FSI effect.
%%%%%%%%%%%%%%%%%%%%%%%%%%%%%%%%%%%%%%%
\subsection{Single-spin asymmetries}
\label{sec51}
\begin{figure}[htp]
\includegraphics[scale=0.73]{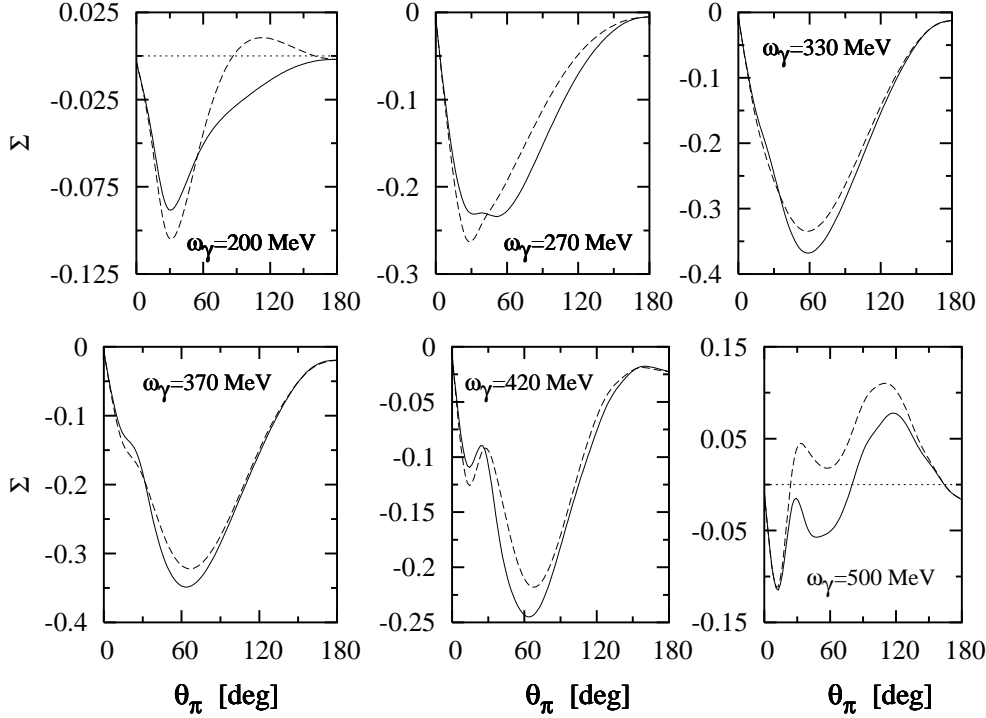}
\caption{Linear photon asymmetry $\Sigma$ for $d(\vec\gamma,\pi^-)pp$ 
  as a function of the emission pion angle $\theta_{\pi}$ in the 
  lab frame for various photon lab-energies $\omega_{\gamma}$. 
  Notation of curves: dashed: IA; solid: IA+$NN$-rescattering.} 
\label{linear}
\end{figure}

\begin{figure}[htp]
\includegraphics[scale=0.75]{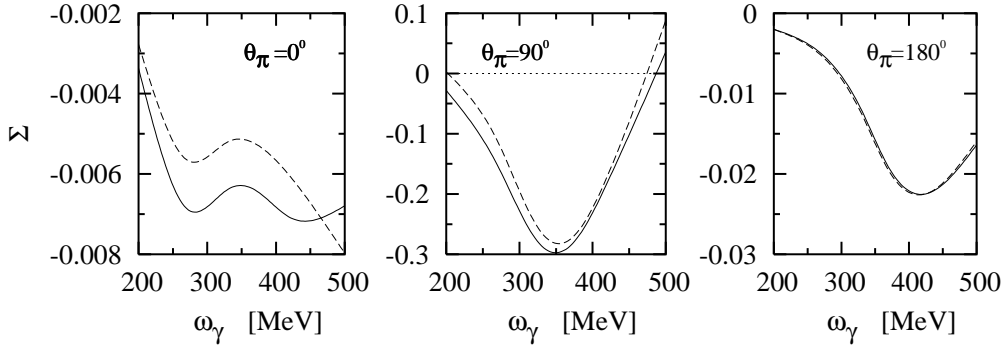}
  \caption{Linear photon asymmetry $\Sigma$ for $d(\vec\gamma,\pi^-)pp$ 
    as a function of $\omega_{\gamma}$ at fixed values of $\theta_{\pi}$. 
    Notation as in Fig.\ \ref{linear}.}
  \label{linearomega}
\end{figure}

\begin{figure}[htp]
\includegraphics[scale=0.73]{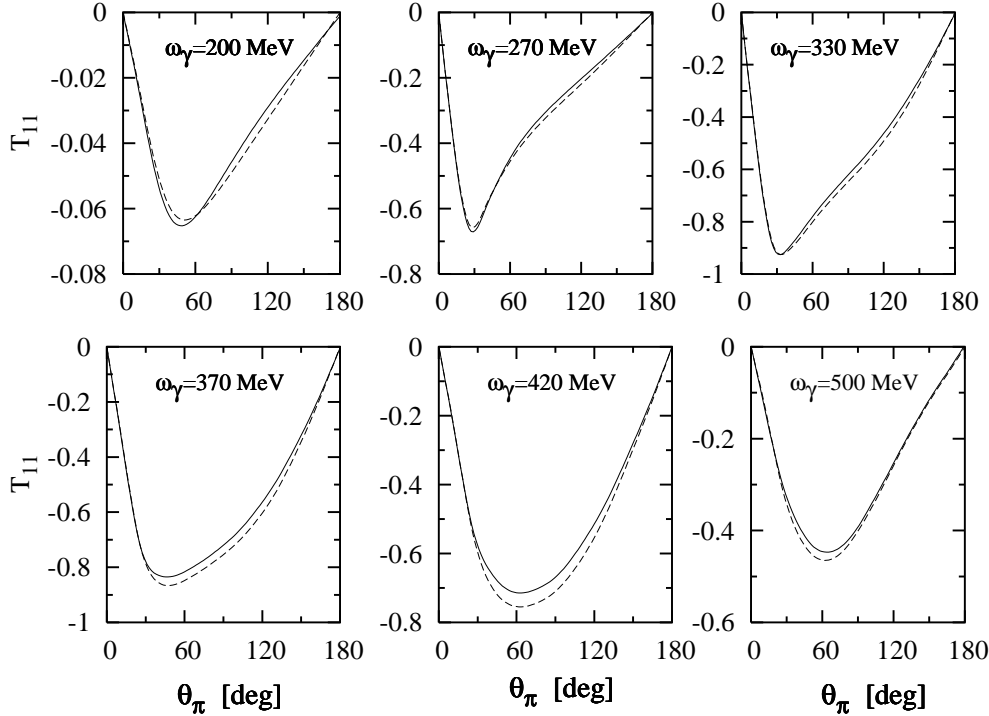}
\caption{Vector target asymmetry $T_{11}$ for $\vec{d}(\gamma,\pi^-)pp$ 
  as a function of $\theta_{\pi}$ for various energies. 
  Notation as in Fig.\ \ref{linear}.} 
\label{t11}
\end{figure}

\begin{figure}[htp]
\includegraphics[scale=0.75]{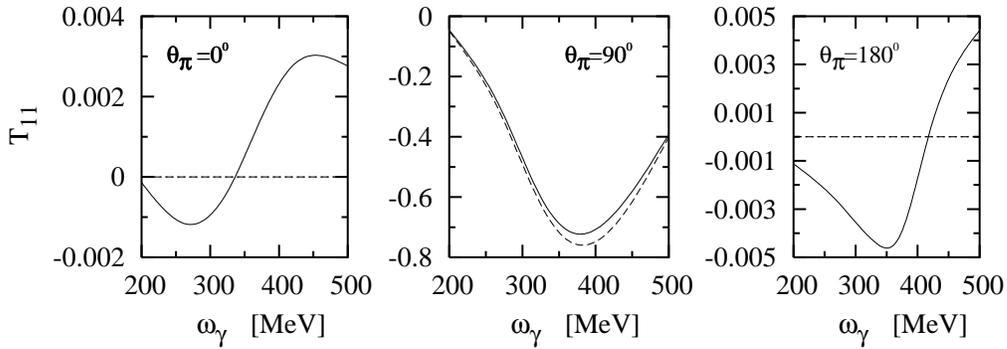}
  \caption{Vector target asymmetry $T_{11}$ for $\vec{d}(\gamma,\pi^-)pp$ 
    as a function of $\omega_{\gamma}$ at fixed values of $\theta_{\pi}$. 
    Notation as in Fig.\ \ref{linear}.}  
  \label{t11omega}
\end{figure}

\begin{figure}[htp]
\includegraphics[scale=0.73]{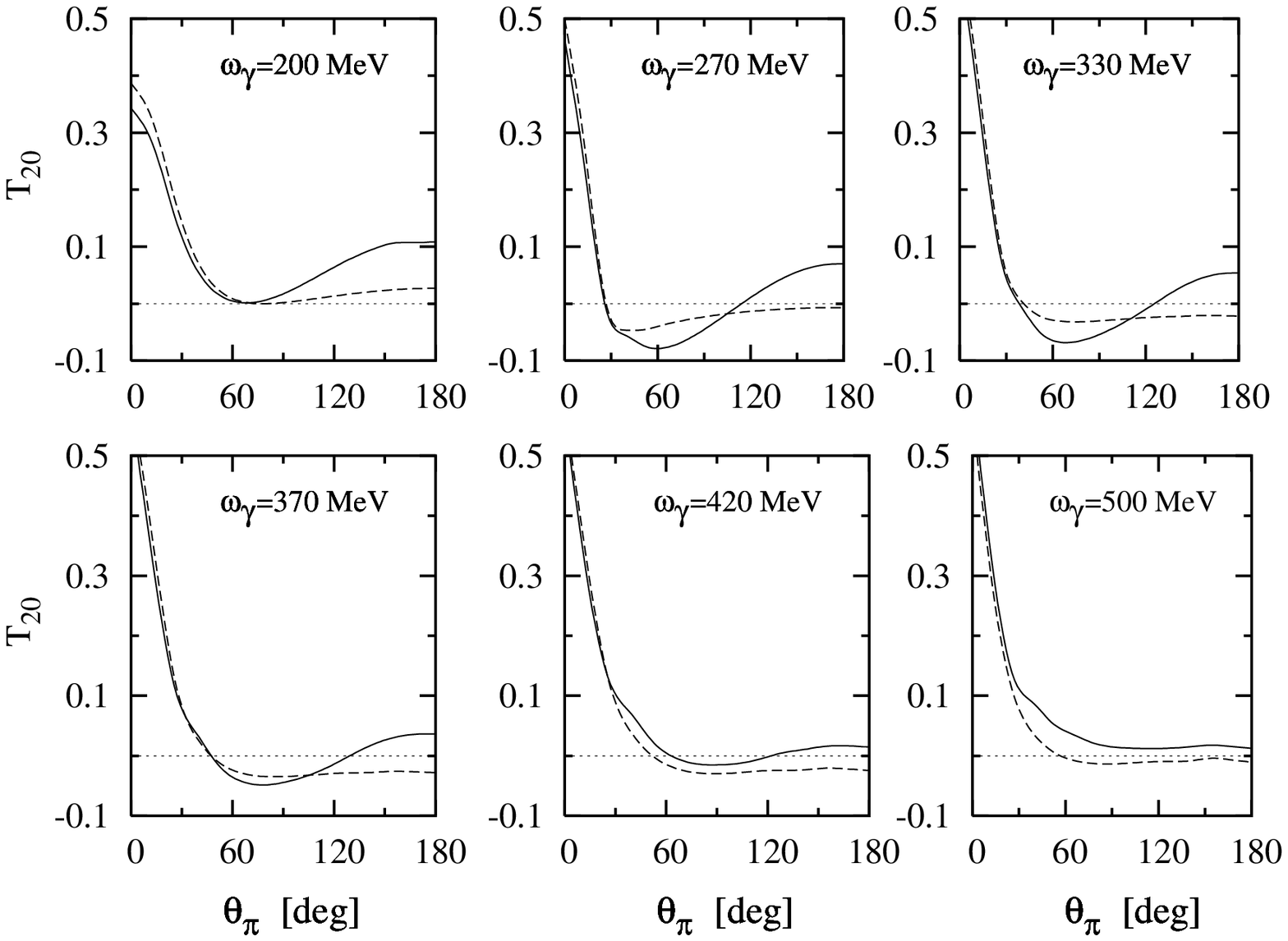}
\caption{Tensor target asymmetry $T_{20}$ for $\vec{d}(\gamma,\pi^-)pp$ 
  as a function of $\theta_{\pi}$ for various energies. 
  Notation as in Fig.\ \ref{linear}.}
  \label{t20}
\end{figure}

\begin{figure}[htp]
\includegraphics[scale=0.75]{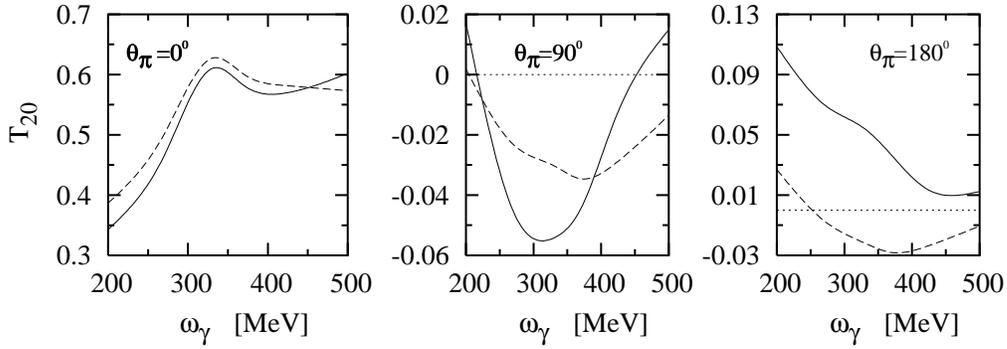}
  \caption{Tensor target asymmetry $T_{20}$ for $\vec{d}(\gamma,\pi^-)pp$ 
    as a function of $\omega_{\gamma}$ at fixed values of $\theta_{\pi}$. 
    Notation as in Fig.\ \ref{linear}.} 
  \label{t20omega}
\end{figure}

\begin{figure}[htp]
\includegraphics[scale=0.73]{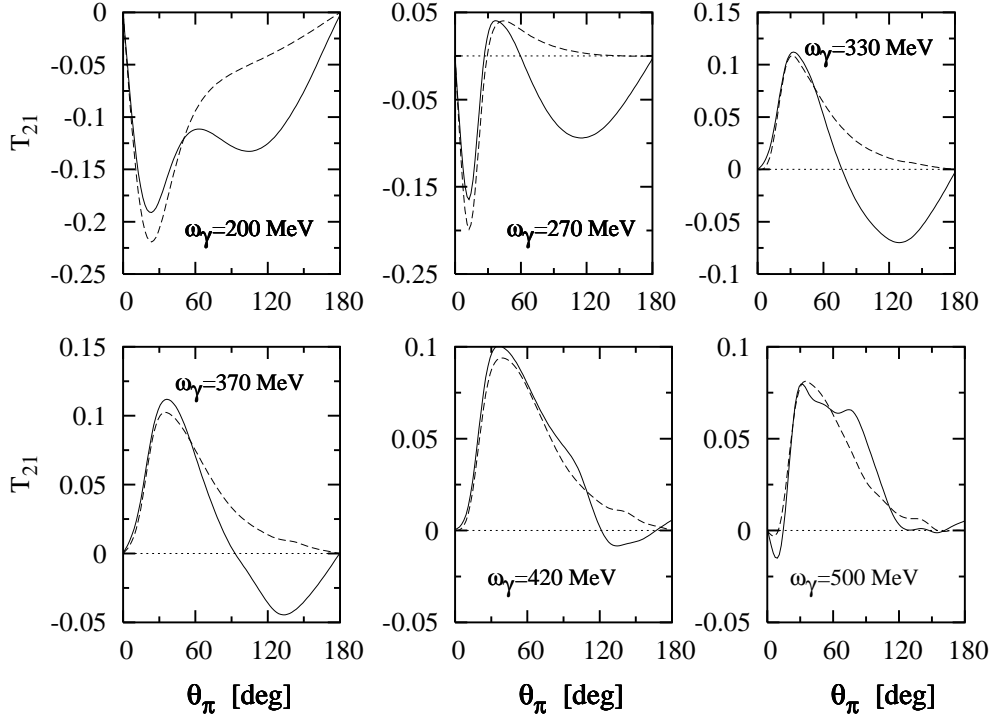}
\caption{Tensor target asymmetry $T_{21}$ for $\vec{d}(\gamma,\pi^-)pp$ 
  as a function of $\theta_{\pi}$ for various energies. 
  Notation as in Fig.\ \ref{linear}.}
  \label{t21}
\end{figure}

\begin{figure}[htp]
\includegraphics[scale=0.75]{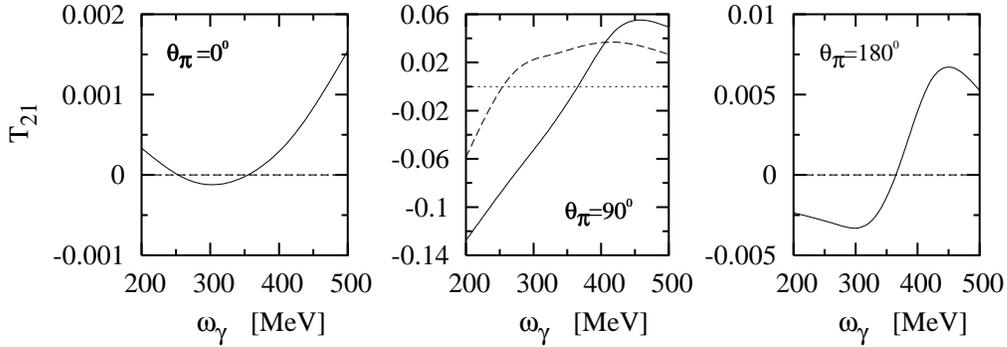}
  \caption{Tensor target asymmetry $T_{21}$ for $\vec{d}(\gamma,\pi^-)pp$ 
    as a function of $\omega_{\gamma}$ at fixed values of $\theta_{\pi}$. 
    Notation as in Fig.\ \ref{linear}.}
  \label{t21omega}
\end{figure}

\begin{figure}[htp]
\includegraphics[scale=0.73]{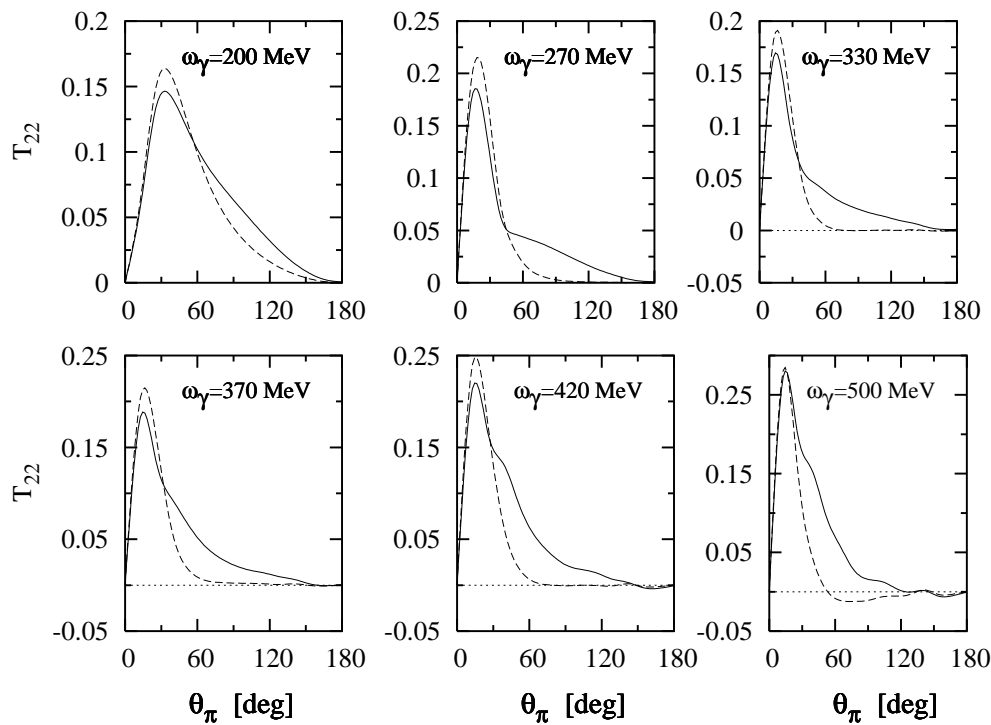}
\caption{Tensor target asymmetry $T_{22}$ for $\vec{d}(\gamma,\pi^-)pp$ 
  as a function of $\theta_{\pi}$ for various energies. 
  Notation as in Fig.\ \ref{linear}.}
  \label{t22}
\end{figure}

\begin{figure}[htp]
\includegraphics[scale=0.75]{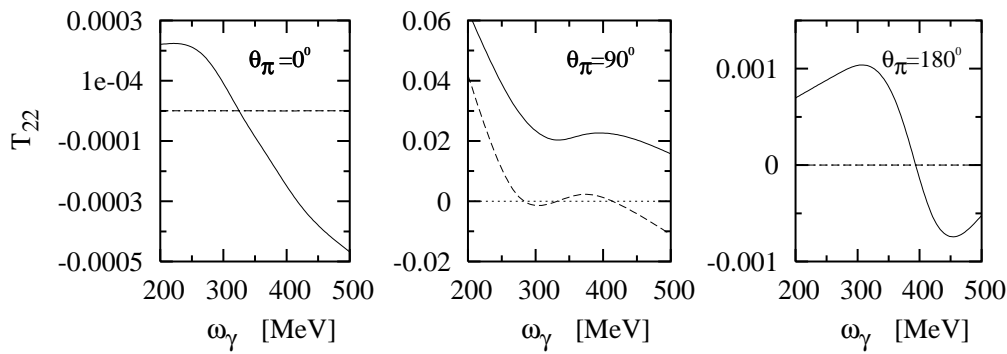}
  \caption{Tensor target asymmetry $T_{22}$ for $\vec{d}(\gamma,\pi^-)pp$ 
    as a function of $\omega_{\gamma}$ at fixed values of $\theta_{\pi}$. 
    Notation as in Fig.\ \ref{linear}.} 
  \label{t22omega}
\end{figure}

%%%%%%%%%%%%%%%%%%%%%%%%%%%%%%%%%%%%%%%
%%%%%%%%%%%%%% Sigma %%%%%%%%%%%%%%%%%%%%
We begin the discussion by presenting our results for the single-spin
asymmetries $\Sigma$, $T_{11}$, $T_{20}$, $T_{21}$, and $T_{22}$ in
the pure IA and with $NN$-FSI, as shown in Figs.\ \ref{linear} through
\ref{t22omega}. The photon asymmetry $\Sigma$ for linearly polarized
photons at various photon lab-energies ($\omega_{\gamma}=200$, $270$,
$330$, $370$, $420$ and $500$ MeV) is plotted in Fig.\ \ref{linear} as
a function of pion angle $\theta_{\pi}$ in the lab frame. 
%%%%%%%%%%%%%%%%%%%%
%%%%%%%%%%%%%%%%%%%%
%%%%%%%%%%%%%%%%%%
%%%%%%%%%%%%%%%%%%
In order to show the sensitivity of the results to the photon
lab-energy $\omega_{\gamma}$, we present in Fig.\ \ref{linearomega}
the asymmetry $\Sigma$ as a function of $\omega_{\gamma}$ for fixed
pion angles ($\theta_{\pi}=0^{\circ}$, $90^{\circ}$ and
$180^{\circ}$). In the photon energy domain of the present work
($\Delta$(1232)-resonance region), the magnetic multipoles dominate
over the electric ones, due to the excitation of the
$\Delta$-resonance. This is clear from the dominately negative values
of $\Sigma$ as shown in Fig.\ \ref{linear}.  We also see that the
asymmetry $\Sigma$ which is a ratio of cross sections is found to be
sensitive to the energy of the incoming photon. 

The left-top and right-bottom panels in Fig.\ \ref{linear} show that
small positive values are found at $\omega_{\gamma}=200$ and $500$
MeV. After including $NN$-FSI, we see that these positive values have
disappeared at $200$ MeV, while at $500$ MeV these values are still
evident at backward pion angles. At extreme forward and backward pion
angles one can see, that the asymmetry $\Sigma$ is relatively small in
comparison to the results when $\theta_{\pi}$ changes from about
$30^{\circ}$ to $120^{\circ}$. One also notices, that the contribution
from $NN$-rescattering is much important in this region, in particular
in the peak position. For lower and higher photon energies, one finds
the strongest effect by $NN$-rescattering. It is also clear from Fig.\ 
\ref{linear} that the asymmetry $\Sigma$ vanishes at
$\theta_{\pi}=0^{\circ}$, whereas a tiny negative value at
$180^{\circ}$ is observed.

%%%%%%%%%%%%%%%% T11 %%%%%%%%%%%%%%%%%%%%%%%%%%%%%%%%%
We show in Fig.\ \ref{t11} the vector target asymmetry $T_{11}$ as a
function of pion angle in the lab frame for various photon
lab-energies. The energy dependence of this asymmetry for fixed pion
angles is displayed in Fig.\ \ref{t11omega}. In general, one can see
that the $T_{11}$ asymmetry has negative values even after including
the $NN$-FSI. These values come mainly from the Born terms.  It is
also noticeable that the $T_{11}$-asymmetry vanish at
$\theta_{\pi}=0^{\circ}$ and $180^{\circ}$ which is not the case for
the linear photon asymmetry $\Sigma$ as shown in Figs.\ \ref{linear}
and \ref{linearomega}. 
%%%%%%%%%%%
%%%%%%%%%%%
%%%%%%%%%%%
%%%%%%%%%%%
It is also apparent that the contribution of
$NN$-FSI - the difference between the dashed and the solid curves - is
very small, almost completely negligible at extreme forward and
backward pion angles. Adding $NN$-FSI gives a slight increase (in
absolute size decrease) of a few percent in the maximum.
%%%%%%%%%%%%%%
%%%%%%%%%%%%%%
%%%%%%%%%%%%%%
%%%%%%%%%%%%%%

%%%%%%%%%%%%%%%%%%%% T_20 %%%%%%%%%%%%%%%%%%%%%%%%%%%%%
Let us discuss now the results for the tensor target asymmetries
$T_{20}$, $T_{21}$ and $T_{22}$ as shown in Figs.\ \ref{t20} through
\ref{t22omega}.  We emphasize that the tensor target asymmetries are
found to be sensitive to the $NN$-FSI. In Figs.\ \ref{t20} and
\ref{t20omega} we depict the results for the asymmetry $T_{20}$ as a
function of pion angle $\theta_{\pi}$ for various energies (Fig.\ 
\ref{t20}) and of photon lab-energy $\omega_{\gamma}$ for fixed pion
angles (Fig.\ \ref{t20omega}). It is known \cite{Wil96} that the
tensor target asymmetry $T_{20}$ serves as a special tool to
disentangle different reaction mechanisms. Comparing with the results
for linear photon and vector target asymmetries we found that the
$T_{20}$-asymmetry has relatively large positive values at pion
forward angles (at $\theta_{\pi}<30^{\circ}$) while small negative
ones are found when $\theta_{\pi}$ changes from $30^{\circ}$ to
$180^{\circ}$. Only at energies above the $\Delta$-region we observe
small negative values at extreme backward angles. 
%%%%%%%%%%%%
%%%%%%%%%%%%
%%%%%%%%%%%%
%%%%%%%%%%%%%%
It is also obvious
that the negative values appeared at extreme backward angles have 
totally disappeared after including the final-state $NN$-rescattering.
Furthermore, quite a significant contribution from $NN$-FSI is found
at pion backward angles. Intuitively, one would always expect that
rescattering mechanisms become more important at higher momentum
transfers, i.e., for larger scattering angles at fixed energy, since
rescattering provides a means to share the momentum transfer between
the two nucleons.

%%%%%%%%%%%%%%%%% T_21 %%%%%%%%%%%%%%%%%%%%%%%%
Predictions for the tensor target asymmetry $T_{21}$ are shown in
Figs.\ \ref{t21} (as a function of $\theta_{\pi}$ for various
$\omega_{\gamma}$) and \ref{t21omega} (as a function of
$\omega_{\gamma}$ for fixed $\theta_{\pi}$). It is clear that the
$T_{21}$ asymmetry has positive values at pion forward angles, which
vanish at $\theta_{\pi}=180^{\circ}$.  At energies lower and higher
than the $\Delta$(1232)-resonance region we see that $T_{21}$ has
negative values at extreme forward angles. Similar to the case in the
vector target asymmetry $T_{11}$, we found that the asymmetry $T_{21}$
vanishes at $\theta_{\pi}=0^{\circ}$ and $\theta_{\pi}=180^{\circ}$.
One readily notes, that the $NN$-rescattering mechanism becomes much
more effective at pion backward angles, in particular at
$\theta_{\pi}\simeq 120^{\circ}$.

%%%%%%%%%%%%%%%%%% T_22 %%%%%%%%%%%%%%%%%%%%%%%%%%%
In Figs.\ \ref{t22} and \ref{t22omega} we present the results for the
tensor target asymmetry $T_{22}$ as a function of $\theta_{\pi}$ for
various $\omega_{\gamma}$ (Fig.\ \ref{t22}) and of $\omega_{\gamma}$
for fixed $\theta_{\pi}$ (Fig.\ \ref{t22omega}). The same values of
photon lab-energies as in the 
%%%%%%%%%%
%%%%%%%%%%
%%%%%%%%%%
%%%%%%%%%
previous figures have been used. Similar
to the results of Figs.\ \ref{t20} through \ref{t21omega}, one can see
that the $T_{22}$ asymmetry is sensitive to the values of pion angle
$\theta_{\pi}$ and photon energy $\omega_{\gamma}$. It is observed
that, the asymmetry $T_{22}$ has qualitatively large positive values
at extreme forward angles. It is also obvious that $T_{22}$ vanishes
at $\theta_{\pi}=0^{\circ}$ and $180^{\circ}$.

From the foregoing discussion for the tensor target asymmetries it is
apparent that the contribution of $NN$-FSI is important and must be
considered in the analysis of the forthcoming experiments. This means,
in particular with respect to a test of theoretical models for
pion-production amplitudes on the neutron, that one needs a reliable
description of the scattering process. Hopefully, these predictions
can be tested in the near future when the data from the on-going
experiments become available.
%%%%%%%%%%%%%%%%%%%%%%%%%%%%%%%%%%%%%%%%%%%%%%%%%%%%%%%%%
\subsection{Beam-target double polarization asymmetries}
\label{sec52}
\begin{figure}[htp]
\includegraphics[scale=0.73]{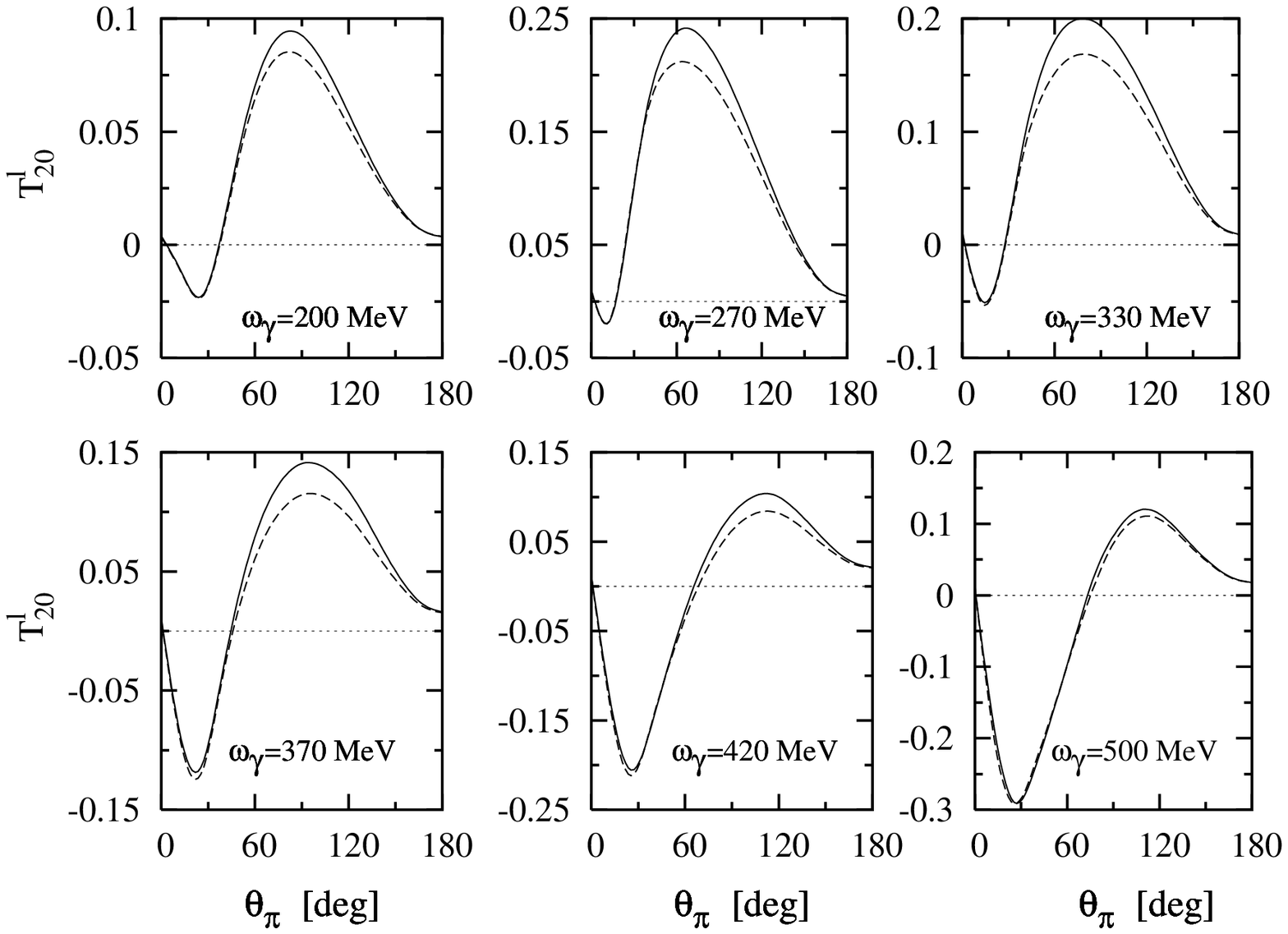}
\caption{The double-polarization asymmetry $T_{20}^{\ell}$ 
  for $\vec{d}(\gamma,\pi^-)pp$ as a function of $\theta_{\pi}$ for various 
  energies. Notation as in Fig.\ \ref{linear}.} 
  \label{t20l}
\end{figure}

\begin{figure}[htp]
\includegraphics[scale=0.75]{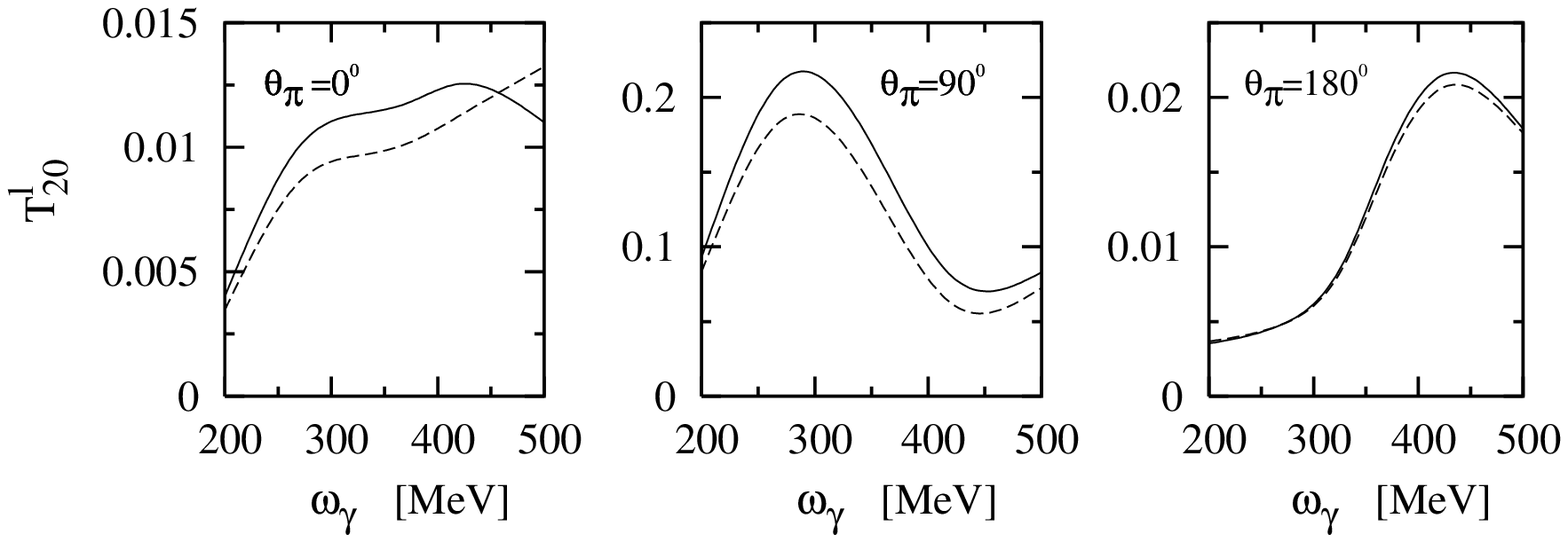}
  \caption{The double-polarization asymmetry
  $T_{20}^{\ell}$ for $\vec{d}(\vec\gamma,\pi^-)pp$ as a function of
  $\omega_{\gamma}$ at fixed values of $\theta_{\pi}$. Notation as in 
  Fig.\ \ref{linear}.} 
  \label{t20lomega}
\end{figure}

\begin{figure}[htp]
\includegraphics[scale=0.73]{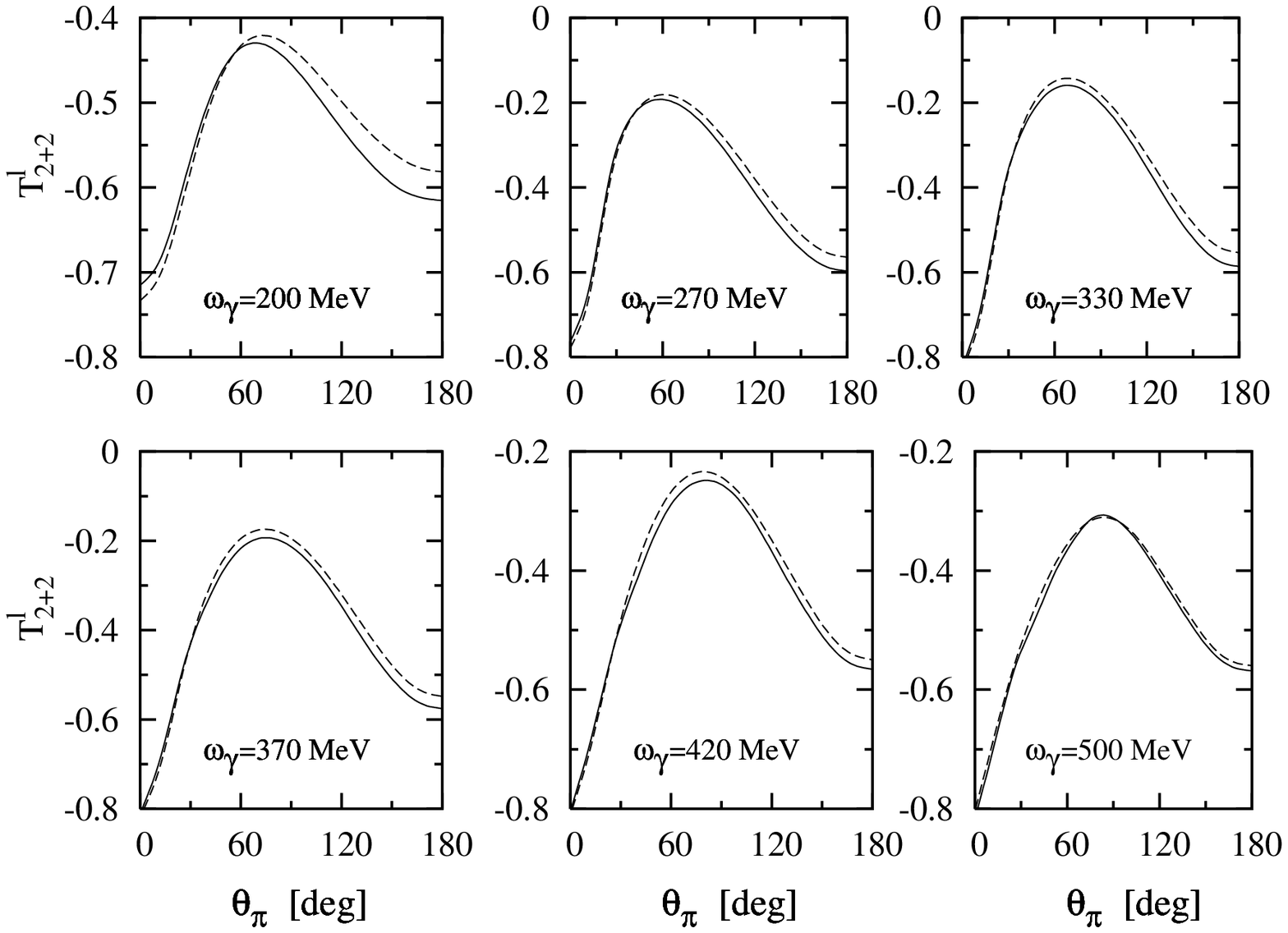}
\caption{The double-polarization asymmetry $T_{2+2}^{\ell}$ 
  for $\vec{d}(\vec\gamma,\pi^-)pp$ as a function of $\theta_{\pi}$ for 
  various energies. Notation as in Fig.\ \ref{linear}.}  
  \label{t2p2l}
\end{figure}

\begin{figure}[htp]
\includegraphics[scale=0.75]{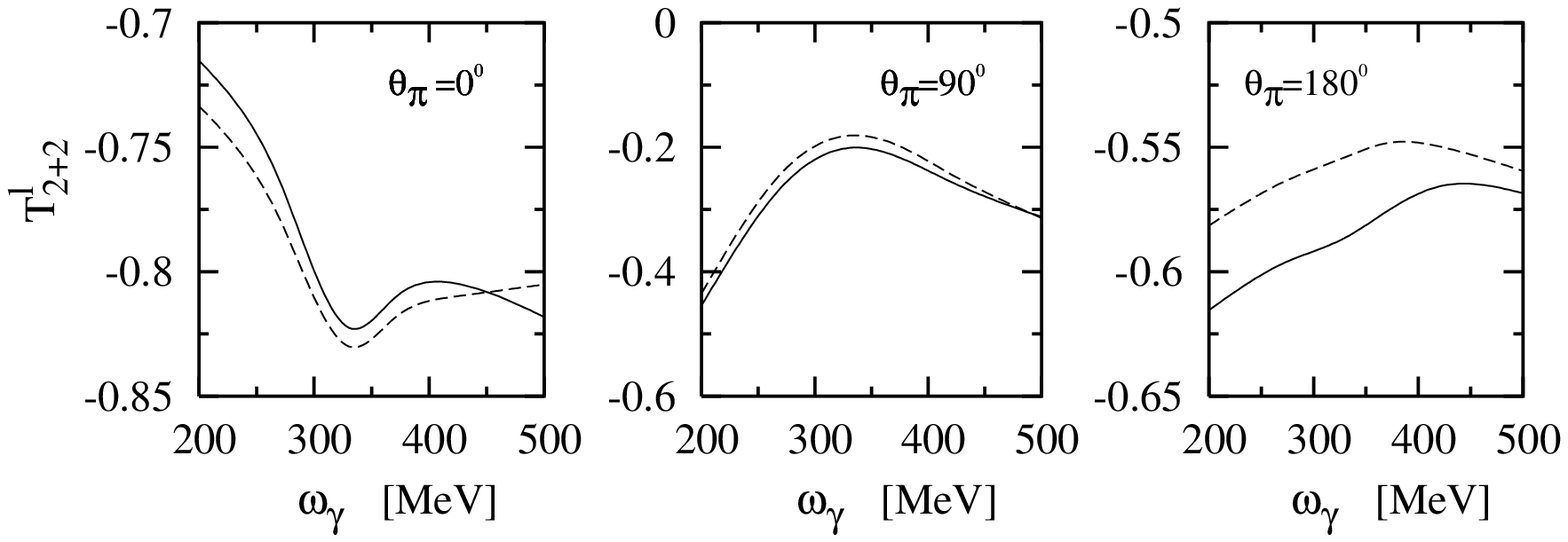}
  \caption{The double-polarization asymmetry $T_{2+2}^{\ell}$ 
    for $\vec{d}(\vec\gamma,\pi^-)pp$ as a function of $\omega_{\gamma}$ 
    at fixed values of $\theta_{\pi}$. Notation as in Fig.\ \ref{linear}.}  
  \label{t2p2lomega}
\end{figure}

%%%%%%%%%%%%%%%%%%%%%%%%%%%%%%%%%%%%%%%%%%%%%%%%%%%%%%%%%
As next we discuss the influence of $NN$-rescattering on beam-target
double-polarization asymmetries as shown in Figs.\ \ref{t20l} through
\ref{t2p2lomega}. Interest in double-polarization observables comes
from the recent technical improvements of electron accelerator
facilities with both polarized beams and polarized targets. In view of
these recent developments, it will soon be possible to measure
double-spin observables with precision.

As already mentioned in section \ref{sec4}, the asymmetries
$T_{10}^c$, $T_{11}^c$, $T_{10}^{\ell}$, $T_{1\pm 1}^{\ell}$,
$T^c_{20}$, $T^c_{21}$, $T^c_{22}$ and $T_{2\pm 2}^{\ell}$ do vanish,
whereas the spin asymmetries $T_{20}^{\ell}$ and $T_{2\pm 2}^{\ell}$
do not. Moreover, we would like to mention that the values for the
$T_{2+2}^{\ell}$ asymmetry are found to be identical with the values
of $T_{2-2}^{\ell}$. Therefore, in what follows we shall discuss the
results for only the $T_{20}^{\ell}$ and $T_{2+2}^{\ell}$ asymmetries.
%%%%%%%%%
%%%%%%%%
%%%%%%%%
%%%%%%%%

We show in Figs.\ \ref{t20l} and \ref{t20lomega} the influence of
$NN$-rescattering on the longitudinal double-spin asymmetry
$T_{20}^{\ell}$ as a function of $\theta_{\pi}$ for various energies
(Fig.\ \ref{t20l}) and of $\omega_{\gamma}$ for fixed angles (Fig.\ 
\ref{t20lomega}). We see that $T_{20}^{\ell}$ has negative values at
forward pion angles around $\theta_{\pi}< 60^{\circ}$ which is not the
case at backward pion angles.  These negative values increase (in
absolute value decrease) with increasing the photon energy.  At
extreme backward angles, we see that $T_{20}^{\ell}$ has small
positive values.
%%%%%%%%
%%%%%%%%
%%%%%%%%
%%%%%%%%
We emphasize that the contribution from $NN$-FSI is much important in
the energy region around the $\Delta$-resonance, especially in the
peak position. One notices that the inclusion of $NN$-FSI leads to an
overestimation by about 10$\%$ of the plane wave results. For lower
and higher energies, one can see that the $NN$-FSI effect is small.

The predictions for the longitudinal double-polarization asymmetry
$T_{2+2}^{\ell}$ are plotted in Figs.\ \ref{t2p2l} and
\ref{t2p2lomega} as a function of pion angle for various energies and
of the photon lab-energy for fixed pion angles. The same values of
photon lab-energies and pion angles as the abovementioned cases are
used. In general, one readily notices that the longitudinal asymmetry
$T_{2+2}^{\ell}$ has negative values even after including the
contribution of $NN$-rescattering. Evidently, the values of
$T_{2+2}^{\ell}$ are sensitive to the photon energy and/or pion angle.
We also emphasize that the $NN$-FSI shows rather insignificant effect.

We would like to mention that we have obtained essentially the same
results for the foregoing polarization asymmetries if we take the Bonn
r-space potential model instead of the Paris one to estimate the
half-off-shell partial wave $NN$ amplitudes in (\ref{eq:10}). 
%%%%%%%%%%%%%%%%%%%%%%%%%%%%%%%%%%%%%%%%%%%%%%%%%%%%%%%%%
\subsection{Comparison with experimental data}
\label{sec53}
\begin{figure}[htp]
  \includegraphics[scale=0.75]{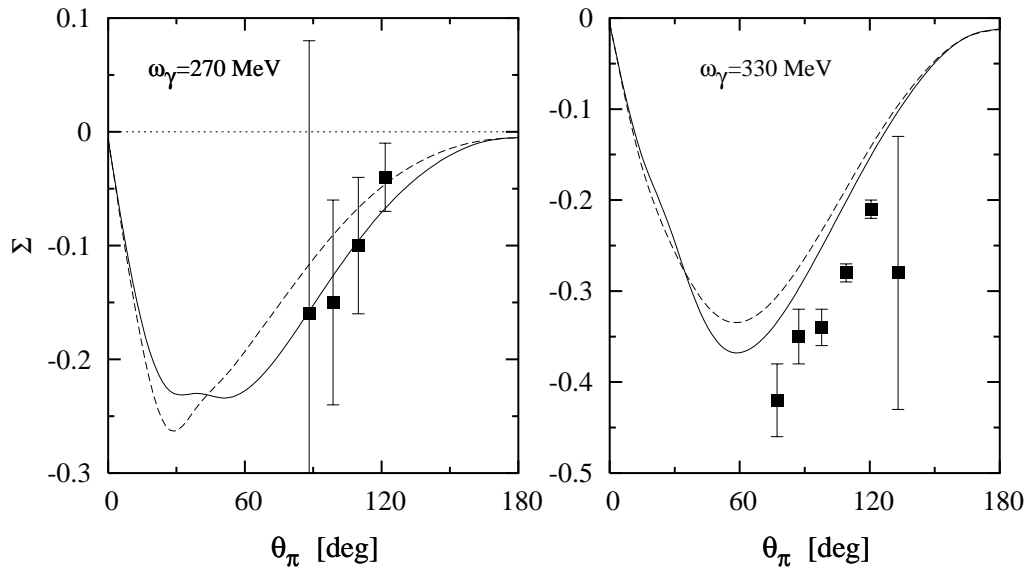}
  \caption{Linear photon asymmetry $\Sigma$ for the reaction
    $d(\vec\gamma,\pi^-)pp$ as a function of $\theta_{\pi}$ at 
    $\omega_{\gamma}=270$ and $330$ MeV in comparison with the 
    preliminary data from LEGS \cite{LEGS}. Notation as in 
    Fig.\ \ref{linear}.}
  \label{linearexp}
\end{figure}

%%%%%%%%%%%%%%%%%%%%%%%%%%%%%%%%%%%%%%%%%%%%%%%%%%%%%%%%%
Here we compare our results with the available experimental data. Most
recently, a few preliminary data points are available from the LEGS
Spin collaboration \cite{LEGS} for the linear photon asymmetry
$\Sigma$, but with regard to the other polarization observables, there
are no data whatsoever. In Fig.\ \ref{linearexp} we confront these
experimental data with our theoretical results in the pure IA (dashed
curves) and with $NN$-rescattering (solid curves) at two photon
lab-energies $\omega_{\gamma}=270$ and $330$ MeV in order to establish
the present status of our knowledge of this important asymmetry. 
We see that the general feature of the data is reproduced. However,
the discrepancy is rather significant in the region where the photon
energy close to the $\Delta$-resonance. This could be due to the
higher order rescattering mechanisms which are neglected in this work.
It is obvious, therefore, that for a conclusive interpretation, one
has to include all corrections of FSI and two-body effects
consistently. In the same figure, we also show the results from the IA
only (dashed curves). It is seen that the $NN$-rescattering yields an
about 10$\%$ effect in the region of the peak position. We found that
this is mainly due to the interference between the IA amplitude and
the $NN$-FSI amplitude.

In agreement with our previous results \cite{Dar04JG}, one notes that
the pure IA (dashed curves in Fig.\ \ref{linearexp}) cannot describe
the experimental data. The inclusion of $NN$-FSI leads at
$\omega_{\gamma}=270$ MeV to a quite satisfactory description of the
data, whereas at $330$ MeV $NN$-FSI effect is small and therefore 
differences between theory and experiment are still evident. An 
experimental check of the $\Sigma$-asymmetry covering a large range
for the emission pion angle and the photon lab-energy would provide an
additional significant test of our present theoretical understanding
of this spin observable.
%%%%%%%%%%%%%%%%%%%%%%%
%%%%%%%%%%%%%%%%%%%%%%%
%%%%%%%%%%%%%%%%%%%%%%%%%%%%%%%%%%%
\section{Conclusions}
\label{sec6}
%%%%%%%%%%%%%%%%%%%%%%%%%%%%%%%%%%%
In the present paper, we have presented calculations for polarization
observables in incoherent negative pion photoproduction from the
deuteron in the $\Delta$(1232)-resonance region taking into account
the $NN$-rescattering in the final state. The influence of $NN$-FSI on
various single- and double-spin observables is investigated.  In view
of the results presented in this paper, one has to conclude that the
contribution of $NN$-rescattering effect must be taken into account in
the analysis of experimental data for spin observables. Compared to
the preliminary experimental data for the linear photon asymmetry
\cite{LEGS}, one notices a remaining discrepancy even if $NN$-FSI is
included.  Therefore, further investigations are required.
Specifically, it will be interesting to see whether this remaining
discrepancy is connected with a complete three-body treatment of the
final $\pi NN$ system.

Concluding this paper, we would like to mention, that the forthcoming
experimental data on spin-dependent observables are very welcome since
it is expected that they might play an important role in deepening our
theoretical understanding of the incoherent pion photoproduction
reaction from the deuteron. On the theoretical side, we have to
improve the treatment by including higher order rescattering and
two-body effects in order to obtain a more realistic description of
these important spin observables. Furthermore, an independent
calculations in the framework of effective field theory would be very
interesting.
%%%%%%%%%%%%%%%%%%%%%%%%%%%%%%%%%%%%%%%%%%%%%%%%%%%%%%%%%%%%%%%%%%%%%%%
\begin{ack}
  This work is supported in part by the Bibliotheca Alexandrina -
  Center for Special Studies and Programs - under grant number:
  2602314 Sohag 2$^{nd}$-Sohag. We are indebted to Profs.\ H.\ 
  Arenh\"ovel, T.-S.\ Harry Lee, T.\ Sato and A.\ Sandorfi for
  fruitful discussions and valuable informations.
\end{ack}
%%%%%%%%%%%%%%%%%%%%%%%%%%%%%%%%%%%%%%%%%%%%%%%%%%%%%%%%%%%%%%%%%%%%%%%%%%

%%%%%%%%%%%%%%%%%%%%%%%%%%%%%%%%%%%%%%%%%%%%%%%%%%%%%%%%%%%%%%%%%%%%%%%%%%

\end{document}